\def\fun#1#2{\lower3.6pt\vbox{\baselineskip0pt\lineskip.9pt
  \ialign{$\mathsurround=0pt#1\hfil##\hfil$\crcr#2\crcr\sim\crcr}}}
\relax
\documentstyle[12pt]{article}
\advance\textheight by \topskip
\begin{document}
\begin{flushright}
NI94002\\
QMW-PH-94-7\\
SU-ITP-94-8\\
hep-th/9404006

\end{flushright}
\vspace{-0.2cm}
\begin{center}
{\large\bf SUPERSYMMETRY, TRACE ANOMALY\\
\vskip 1 cm
AND NAKED SINGULARITIES}\\
\vskip 2 cm
{\bf Renata Kallosh\footnote{Permanent address:
Physics Department, Stanford University, Stanford   CA 94305, USA.\\
E-mail address: {\tt kallosh@physics.stanford.edu}}
and Tom\'as Ort\'{\i}n}\footnote{Permanent address: Department of
Physics, Queen Mary and Westfield College, Mile End Road, London E1
4NS, U.K.\\
E-mail address: {\tt ortin@qmchep.cern.ch}}
\vskip 0.05cm
Isaac Newton Institute for Mathematical Sciences,\\
\vskip 0.4 cm
University of Cambridge, 20 Clarkson Road, Cambridge, CB3 0EH, U.K.
\vskip 0.7 cm
\end{center}
\vskip 1.5 cm
\centerline{\bf ABSTRACT}
\begin{quotation}

We discuss stationary supersymmetric bosonic configurations of the
Einstein-Maxwell theory embedded in $N=2$ supergravity.  Some of these
configurations, including the Kerr-Newman solutions with $m = |q|$ and
arbitrary angular momentum per unit mass $a$, exhibit naked
singularities.  However, $N=2$ supergravity has trace anomaly.  The
nonvanishing anomalous energy-momentum tensor of these Kerr-Newman
solutions violates a consistency condition for a configuration to admit
unbroken supersymmetry.  Thus, the trace anomaly of this theory prevents
the supersymmetric solutions from exhibiting naked singularities.

\end{quotation}


\newpage

\section{Introduction}

In this paper we would like to continue our analysis of the relation
between supersymmetry and cosmic censorship, which we started in
\cite{US}.  We have observed that the parameters of the static dilaton
black holes \cite{GM} considered as bosonic solutions of $N=4$
supergravity are constrained due to the existence of supersymmetric
positivity bounds.  The effect of imposing these supersymmetric bounds
on the parameters of black hole solutions is the same as imposing cosmic
censorship: they prevent the solutions from exhibiting naked
singularities.  Based on this example which generalizes the
Reissner-Nordstr\"om black hole case considered in the framework of
$N=2$ supergravity \cite{GH} we conjectured that, in general,
supersymmetry may act as a cosmic censor for static configurations in
asymptotically flat spaces.

The generic feature of theories with global supersymmetry (without
gravity) is the fact that the energy is non-negative, since the
Hamiltonian is a square of supersymmetry charges \cite{Z}.  It looks
plausible that the cosmic censorship role of local supersymmetry is the
generalization of the role of global supersymmetry as warrant of the
positivity of energy in supersymmetric non-gravitational theories.

It may also happen that supersymmetry will help us in justifying the
cosmic censorship hypothesis for certain nonsupersymmetric theories,
just as it happened with the proof of the positivity of energy in
General Relativity.  In that case it was enough to know that this theory
can be consistently {\it embedded} into supergravity \cite{DT}.

In our previous work we have investigated only static (non-rotating)
black holes.  The supersymmetric positivity bound of the
Einstein-Maxwell theory embedded in $N=2$ ungauged supergravity implies
$m^2 \geq q^2$ \cite{GH}, \cite{US}, which guarantees that the static
candidates to end-points of black hole evaporation (i.e. the
Reissner-Nordstr\"om solutions) have an event horizon covering the
singularity.

In the stationary case, though, it was not quite clear whether one could
derive the analogous bound $m^2 \geq a^2 + q^2$ for rotating Kerr-Newman
(KN) black holes from supersymmetry alone.  (Here $a$ is angular
momentum per unit mass.)  Moreover, it was proven by Tod \cite{T} that
all the KN solutions with $m^2 = q^2$ admit Killing spinors.  The KN
black hole is a configuration with $m^2 - a^2 - q^2\geq 0$.  The extreme
one has $ m^2 - a^2 - q^2= 0$.  Any configuration with $m^2 = q^2$ and
non-vanishing angular momentum is far below extremality, which means
that the singularity is not covered by any event horizon.

In fact Tod proved that a whole class of stationary metrics including
Israel-Wilson-Perjes metrics \cite{IW} admit $N=2$ supergravity Killing
spinors.  These solutions have been shown by Hartle and Hawking
\cite{HH} to have, in general, naked singularities.  Therefore in
\cite{US} we restricted our conjecture about supersymmetry as the cosmic
censor only to static (and not stationary) asymptotically flat
solutions.

Note, however, that the appearance of a naked singularity at $m^2 < a^2+
q^2$ is a very subtle effect.  At $m^2 = a^2+ q^2$ the singularity is
deeply hidden under the horizon.  An infinitesimally small decrease of
mass (or increment of angular momentum) immediately destroys the horizon
and makes the singularity naked.  In situations in which small causes
may have large effects, quantum corrections may be very important.

In particular, all supersymmetric KN black holes with a given charge
have the same mass, $m= |q|$, independently of their angular momentum.
In other words, they correspond to degenerate energy eigenstates.  This
degeneracy, being a consequence of supersymmetry, can sometimes be
removed by quantum effects.  And indeed, as we will see, with an account
taken of the trace anomaly, only the state with $a=0$ (the nonrotating
Reissner-Nordstr\"om black hole) remains supersymmetric.

\section{Supersymmetry of Israel-Wilson-Perjes metrics}

We will start by rederiving Tod's result using the standard language of
field theory rather than Newman-Penrose spinor language.  The first
analysis of supersymmetric configurations of $N=2$ supergravity was
performed by Gibbons and Hull in \cite{GH} using the standard field
theory spinors.  They found that the static Papapetrou-Majumdar (PM)
metrics, and in particular, the extremal Reissner-Nordstr\"om black hole
metrics are supersymmetric\footnote{We will not consider pp-wave spaces
in this paper.}.  Later Tod \cite{T} found that, in addition to these
configurations, some other configurations, in particular, some
stationary metrics, also admit supercovariantly constant spinors.  The
class of such metrics admitting $N=2$ Killing spinors is known in
General Relativity \cite{KRAM} as the class of conformal-stationary
Einstein-Maxwell fields with conformally flat 3-dimensional space, or
Israel-Wilson-Perjes (IWP) metrics.  The PM metrics are just the static
IWP metrics.

The IWP metrics and the corresponding electromagnetic fields can be
completely described\footnote{Our notation are given in \cite{US}, and
\cite{TOdual}.  In particular, hatted indices are the curved space
ones.} in terms of a time-independent complex function $V$:
\begin{eqnarray}
&&ds^2 = (V\bar V ) (dt + {\vec w} d{\vec  x})^2 - (V\bar V )^{-1}\,
(d{\vec  x} )^2 \ ,  \quad
\nonumber \\
\nonumber \\
&&{\vec  \nabla} \times {\vec w} = - i (V \bar {V})^{-1}
{\vec \nabla} \log\; (V / \bar V)\, ,
\nonumber \\
\nonumber \\
&&F_{0i} = E_i = {\textstyle\frac{1}{2}}\;
\partial_{\hat \imath} \; (V+\bar V)\, ,
\nonumber \\
\nonumber \\
&&{}^*F_{0i} = iB_i = {\textstyle\frac{1}{2}}\;
\partial_{\hat \imath} \; (V - \bar V)\, .
\label{IW}
\end{eqnarray}

This configuration will be a solution of the Einstein-Maxwell equations
of motion in absence of matter if the complex function $V$ is chosen to
be the inverse of a harmonic function:
\begin{equation}
\triangle  V^{-1} = 0  \ , \qquad  \qquad \triangle  \bar V ^{-1}= 0
\ ,
\label{triangle}
\end{equation}
where $\triangle$ is the flat-space Laplacian in ${\vec x}$.

For real $V$ these configurations are stationary and correspond, as we
have said, to the PM solutions \cite{KRAM}.  These are the only regular
black hole solutions in the IWP class.  All solutions with complex $V$
have naked singularities, according to Hartle and Hawking \cite{HH}.  In
particular, the solution presented in eqs.  (\ref{IW}) includes the
charged KN solution with arbitrary angular momentum and charge equal to
its mass, $m^2=q^2$.  The KN charged rotating black hole solution is
given by
\begin{eqnarray}
ds^2 &=& \left(1- {2mr - q^2\over r^2 + a^2 \cos^2 \theta}\right)dt^2
-  {2a\sin^2 \theta\, (2mr -q^2)\over r^2 + a^2 \cos^2 \theta}dtd\phi
\nonumber\\
&-& (r^2 + a^2 \cos^2 \theta )\left({dr^2 \over  r^2 + a^2 + q^2 -
2mr} + d\theta^2\right )
\nonumber\\
& -& \sin^2\theta\left( r^2 + a^2 + {a\sin^2 \theta\, (2mr -q^2)\over
r^2 + a^2 \cos^2 \theta} (2mr - q^2) \right) d\phi^2 \ .
\label{KN}
\end{eqnarray}
When $m^2=q^2$ and $a$ is arbitrary, this metric can be brought to the
form of eq.  (\ref{IW}) \cite{IW}.  In Cartesian coordinates the complex
harmonic function is
\begin{equation}
V= 1 + {m\over \sqrt  {x^2+y^2 +(z-ia)^2} } \ .
\label{V}
\end{equation}
In terms of more suitable oblate spheroidal coordinates $x+iy = [(r-m)^2
+ a^2]^{1/2}\sin \theta e^{i\phi}, \; z = (r-m) \cos \theta$, the
function $V$ takes the form
\begin{equation}
V= 1 + {m\over  r- m - i a \cos\theta }\, ,
\label{R}
\end{equation}
and
\begin{equation}
V\bar V = {(r-m)^2- a^2 \cos^2\theta \over r^2 + a^2
\cos^{2}\theta}\, ,
\label{VV}
\end{equation}
so the Euclidean 3-metric becomes
\begin{equation}
(d{\vec x})^2 =\Bigl((r-m)^2 + a^2 \cos^2 \theta\Bigr)\left({dr^2
\over (r-m)^2 + a^2 } + d \theta^2\right) + \Bigl((r-m)^2 + a^2\Bigr)
\sin^2 \theta d\phi^2 \ .
\label{dx}
\end{equation}
The corresponding $\vec{\omega}$ is
\begin{equation}
{\vec w}\cdot d{\vec  x}= {(2mr -m^2) a \sin^2 \theta \over (r-m)^2 +
a^2 \cos^2 \theta } d\phi \ .
\label{w}
\end{equation}
Substituting eqs.  (\ref{VV}), (\ref{dx}) and (\ref{w}) into eq.
(\ref{IW}) and comparing the result with eq.  (\ref{KN}) one can see
that this particular IWP metric with function $V$ given by eq.
(\ref{V}) coincides with that of a KN charged rotating black hole with
$m^2=q^2$.

With this in mind we can analyse the problem of supersymmetry for the
general class of metrics (\ref{IW}), since the KN solution with
$m^2=q^2$ is a particular case of such solutions.  We will come back to
this specific solution when analysing the contribution of the trace
anomaly to the equations of motion.

Consider now the supersymmetry transformation of the gravitino field in
$N=2$ supergravity:
\begin{equation}
{\textstyle\frac{1}{2}} \delta_{\epsilon} \Psi_{\mu I} = \nabla _\mu
\epsilon_I - {\textstyle\frac{1}{2}}\epsilon_{IJ}
\sigma^{ab}F_{ab} \gamma_\mu  \epsilon^J \ , \qquad    I,J = 1,2.
\end{equation}

We want to find time-independent {\it Killing spinors} $\epsilon^{I}$,
i.e. spinors for which the above expression vanishes,
\begin{equation}
\delta_{\epsilon} \Psi_{\mu I} =0 \ ,
\end{equation}
and whose partial time derivative vanishes too,
\begin{equation}
\partial_{\hat{0}}\epsilon_{I}=0 \ .
\label{time}
\end{equation}

We already know from Tod's work that these equations will have
nontrivial solutions for IWP metrics, and so we will substitute eqs.
(\ref{IW}) in it, and we will look for the Killing spinors which we know
to exist.  But we will not require the field configurations to satisfy
any specific equations of motion like the Einstein-Maxwell equations of
motion in absence of matter or any other equations.  Thus $V$ will not
be constrained to be the inverse of a harmonic function as in eq.
(\ref{triangle}) and will remain arbitrary for most of our discussion.

It is convenient to express the supersymmetry transformation of the
gravitino in terms of Dirac spinors $\epsilon=\epsilon^{1}+\epsilon_{2}$
and $\psi_{\mu}=\psi_{\mu}^{1}+\psi_{\mu2}$,
\begin{equation}
{\textstyle\frac{1}{2}}\delta_{\epsilon}\psi_{\mu}=
\nabla_{\mu}\epsilon+{\textstyle\frac{1}{2}}\sigma^{ab}F_{ab}
\gamma_{\mu}\gamma_{5}\epsilon\, .
\end{equation}
If we express the chiral Majorana spinors $\epsilon_{I}$ in terms of
two-component Weyl spinors $\tilde{\epsilon}_{I}$ according to our
conventions we have
\begin{equation}
\epsilon_{I}=
\left(
\begin{array}{c}
\tilde{\epsilon}_{I}\\
0
\end{array}
\right)
\, ,
\hspace{1cm}
\epsilon^{I}=
\left(
\begin{array}{c}
0 \\
\tilde{\epsilon}^{I}
\end{array}
\right)
\, ,
\hspace{1cm}
\epsilon=
\left(
\begin{array}{c}
\tilde{\epsilon}_{2} \\
\tilde{\epsilon}^{1}
\end{array}
\right)
\, .
\end{equation}
First we take the time component of the Killing equation
$\delta_{\epsilon} \Psi_{\hat{0} I} =0$.  Using the time-independence of
the spinors we are looking for, eq.  (\ref{time}), we arrive to
\begin{equation}
\left(
\begin{array}{c}
\sigma^{i} [\omega_{0}^{+0i}\tilde{\epsilon}_{2}-
i F^{+0i}\tilde{\epsilon}^{1}] \\
\\
\sigma^{i} [-\omega_{0}^{+0i}\tilde{\epsilon}^{1}-
i F^{+0i}\tilde{\epsilon}_{2}]
\end{array}
\right)=0\, ,
\end{equation}
which, upon use of eqs. (\ref{IW}), implies the following relation
between the Killing spinors:
\begin{equation}
\tilde{\epsilon}^{1} = - i\left({\overline{V}/V}\right)^{\frac{1}{2}}
\tilde{\epsilon}_{2} \, ,
\label{relation1}
\end{equation}
or, in terms of the chiral Majorana spinors
\begin{equation}
\epsilon^{1}+(\overline{V}/V)^{\frac{1}{2}}\gamma^{0}\epsilon_{2}=0\,  .
\label{relation2}
\end{equation}
Now we take the spatial components of the Killing equation
$\delta_{\epsilon} \Psi_{i}=0$.  Using eqs.  (\ref{IW}) and the relation
between the spinors eq.  (\ref{relation1}) we get the following two
equations:
\begin{eqnarray}
\partial_{\hat{\imath}}(\overline{V}^{\frac{1}{2}}\tilde{\epsilon}^{1})
& = & 0\; ,
\nonumber \\
\partial_{\hat{\imath}}(V^{\frac{1}{2}}\tilde{\epsilon}_{2}) & = &0\; ,
\end{eqnarray}
which imply for the chiral Majorana spinors,
\begin{eqnarray}
\epsilon^{1} & = & \overline{V}^{\frac{1}{2}}\epsilon_{(0)}{}^{1}\; ,
\nonumber \\
\epsilon_{2} & = & V^{\frac{1}{2}}\epsilon_{(0) 2}\; ,
\end{eqnarray}
where $\epsilon_{(0)}{}^{1}$ and $\epsilon_{(0) 2}$ are constant chiral
Majorana spinors.  These equations will be consistent with eq.
(\ref{relation2}) if the constant spinors themselves satisfy
\begin{equation}
\epsilon_{(0)}{}^{1} + \gamma^{0} \epsilon_{(0) 2} = 0\ .
\end{equation}

Let us stress that the fundamental difference between supersymmetric
configurations with naked singularities and without them among the IWP
class is the presence or absence of imaginary part in the function $V$.
This is the only function in our Ansatz, which solves Killing spinor
equations and allowed Tod to find supersymmetric configurations without
reference to any equation of motion.

\section{Consistency condition for unbroken supersymmetry}

Consider the classical Einstein-Maxwell action
\begin{equation}
S_{EM}= - {1 \over 4}\int d^4 x\sqrt{-g}\; (R+F^2)\, ,
\end{equation}
which is the bosonic sector of $N=2$ supergravity.  The effective
equations of motion are
\begin{equation}
\frac{\delta S_{EM}}{\delta g^{\mu\nu}}  =  J_{\mu\nu}\, , \qquad
\frac{\delta S_{EM}}{\delta A_{\mu}}     =   J^{\mu} \, .
\label{eqmo}
\end{equation}
The two tensors $J_{\mu\nu}$ and $J^{\mu}$ are the ``right-hand side" of
the metric and electromagnetic vector potential equations of motion.
These two tensors vanish for classical (on-shell) configurations but we
are going to consider general configurations obeying the equations of
motion with $J_{\mu\nu}$ and $J^{\mu}$ nonvanishing in general.  The
notation emphasizes the fact that $J_{\mu\nu}$ is different from the
classical electromagnetic energy-momentum tensor that appears in the
Einstein-Maxwell theory.  Later on we will be interested in
semiclassical configurations for which these tensors are induced by
quantum corrections.

In \cite{KO} we have derived some consistency conditions (Killing Spinor
Identities) that any supersymmetric configuration has to satisfy.  We
are going to show that the only configurations which satisfy these
identities are those with $V$ real.

To find the $N=2$ supergravity Killing Spinor Identities we need  the
function
\begin{equation}
\Omega \equiv \sum_b  J_b \delta_{\epsilon}\,  \phi^b = J_{\mu\nu}
\delta_{\epsilon}\, g^{\mu\nu}+J^{\mu}\delta_{\epsilon}\, A_{\mu}\, ,
\end{equation}
where the supersymmetry transformation of the metric is denoted by
$\delta_{\epsilon} g^{\mu\nu}$, and that of the vector field by
$\delta_{\epsilon} \, A_{\mu}$.  Now we have to differentiate this
function over the gravitino field, and the result has to vanish when
$\epsilon^{I}$ is a Killing spinor.  From now on we will assume this to
be so.  Then the Killing Spinor Identities take the form
\begin{equation}
J^{\mu\nu}\overline{\epsilon}^{I}\gamma_{\nu} +  \frac{1}{2}\;
J^{\mu}\overline{\epsilon}_{\;J}\; \epsilon^{JI} = 0 \ .
\label{example}
\end{equation}
This equation was derived from supersymmetry and therefore the spinor in
this equation is anticommuting.  However, the identity must hold for
commuting spinors as well.  Using commuting spinors it is simple to
derive the consequences of the Killing Spinor Identities for IWP
configurations.  Using the algebraic relation (\ref{relation2}), which
is valid also for commuting Killing spinors, one can derive the
following relation between the function $V$ and the bilinear
combinations of commuting Killing spinors:
\begin{equation}
\bar \epsilon^{I} \gamma_a \epsilon_{I} = (2i\;|V| \, ,
\quad \vec 0) \, ,
\qquad \bar \epsilon_{I} \epsilon_{J} \epsilon^{IJ} = -2i\; V \, .
\end{equation}
Now we may consider eq.  (\ref{example}), where the spinor is commuting.
We multiply this equation by the commuting spinor $\epsilon_{I}$, sum
over the index $I$, and we get for the IWP metrics \begin{equation}
J^{\mu 0}|V|- \frac{1}{2}\; J^{\mu} V = 0\, , \end{equation} which
implies, for complex $V$, $J^{\mu 0}=J^{\mu}=0$.  We are left with
\begin{equation}
J^{\mu\nu}\overline{\epsilon}^{I}\gamma_{\nu}=0\, .
\label{reduced}
\end{equation}
Now we can multiply this equation by a spinor $\eta_{I}$ such that
$\overline{\epsilon}^{I}\gamma_{\nu}\eta_{I}\equiv p_{\nu}\neq 0$.
This gives
\begin{equation}
J^{ij}p_{i}=0\, ,
\end{equation}
which means that
\begin{equation}
J^{ij}=(\eta^{ij}-\frac{p_{i}p_{j}}{p^{2}})f\, .
\label{Jp}
\end{equation}
Finally, if we multiply eq.  (\ref{reduced}) by $\gamma_{\nu} \eta^{J}
\epsilon_{IJ}$ and take into account that $J^{\mu\nu}$ is a symmetric
tensor, we get
\begin{equation}
J^{\mu\nu}\overline{\epsilon}^{I}\gamma_{\nu}\gamma_{\mu}\eta^{J}
\epsilon_{IJ}=
J^{\mu\nu}g_{\mu\nu}\overline{\epsilon}^{I}\eta^{J}
\epsilon_{IJ}=0\, .
\end{equation}
Since $\overline{\epsilon}^{I}\eta^{J}\epsilon_{IJ}\neq 0$, this implies
that $J_{\mu}{}^{\nu}$.  This fact, together with eq.  (\ref{Jp}) proves
that, if $V$ is complex, $J^{\mu\nu}=J^{\nu}=0$.

Thus, for configurations with $V\neq \bar V$, which in general have
naked singularities, the consistency conditions for supersymmetry lead
to relations between the energy-momentum tensor and the Maxwell current
which include a complex function $(V/\overline{V})^{\frac{1}{2}}$.  This
is not acceptable, and the consequence is that the right-hand sides of
the Einstein and Maxwell equations have to vanish for supersymmetric
configurations with {\it complex} $V$ \footnote{Observe that purely {\it
classical} KN configurations have vanishing $J_{\mu \nu}$ and $J^{\mu}$.
Therefore, from the purely classical point of view they {\it are}
supersymmetric.}:
\begin{equation}
J_{\mu \nu} =J^{\mu} = 0 \ .
\end{equation}
In particular we have to require the absence of quantum corrections to
the right-hand side of the trace of Einstein equation for supersymmetric
configurations with complex $V$:
\begin{equation}
R = g^{\mu\nu} J_{\mu\nu} = 0\, .
\end{equation}

\section{The trace anomaly}

The trace anomaly (also called Weyl anomaly) in gravitational
four-dimensional theories was discovered by Capper and Duff about twenty
years ago \cite{CD}.  The existence of this anomaly means that the
conformal invariance under Weyl rescaling of classical gravitational
field syste ms does not survive in the quantum theory.

The trace anomaly of the one-loop on-shell supergravity is given by the
following expression \cite{CD}:
\begin{equation}
T=  g^{\mu\nu} < T_{\mu\nu}> \  = \, {A\over 32 \pi^2} {}^*
R_{\mu\nu\lambda\delta} {}^{*}R^{\mu\nu\lambda\delta} \ .
\end{equation}
The coefficient $A$ is known for all fields interacting with gravity.

The integrated form of the anomaly in Euclidean space expresses the
trace of the energy-momentum tensor through the Euler number of the
manifold,
\begin{equation}
\int d^4x \sqrt{-g} \;T = A \, \chi \ \ .
\end{equation}

The fields of $N=2$ supergravity include a graviton, 2 types of
gravitino and a vector field.  As we see in the table, the anomaly
coefficient $A= {11\over 12}$ of pure $N=2$ supergravity does not
vanish.

\begin{center}
\begin{tabular}{|c|c|c|c|c|c|c|}
\hline \hline
 Field & 360A & $ N=2$ & $ N=2$ & $N=2$& $ N=4$ & $ N=4$  \\
 {}~ & ~ & supergravity  & Yang-Mills &  hypermultiplet &
supergravity &
Yang-Mills
\\
\hline
$ e_{\mu}^a$&848&1&0& 0&1&0\\
\hline
$\psi_{\mu}$&-233&2&0&0&4&0\\
\hline
$A_{\mu}$&-52&1&1&0&6&1\\
\hline
$\chi$&7&0&2&2&4&4\\
\hline
$\phi$&4&0&2&4&1&6\\
\hline
$\phi_{\mu\nu}$&364&0&0&0&1&0\\
\hline
{}~&~&A=${11/12}$ & A=$-{1/12}$ & A=${1/12}$& A=$0$&A=$0$\\
\hline \hline
\end{tabular}
\vskip 0.3cm
Table 1: {Anomalies in N=2 and N=4 supermultiplets.}
\end{center}

The function ${}^*{}R_{\mu\nu\lambda\delta}
{}^{*}R^{\mu\nu\lambda\delta}$ does not vanish in general for IWP
configurations.  In particular, one can calculate this function for the
charged KN solution and check that for arbitrary angular momentum and
charge equal to its mass $m^2=q^2$ this function does not vanish.  One
can use for this purpose the values of non-vanishing components of the
Weyl tensor $C_{abcd}$ and Maxwell tensor $F_{ab}$ given for this
solution in \cite{KRAM} in an isotropic tetrad basis.  The expression
for the anomaly is given by
\begin{equation}
{}^*R_{\mu\nu\lambda\delta}
{}^*R^{\mu\nu\lambda\delta}= 24 (\Psi_2 \Psi_2 + h.c.) - 32 \;(\Phi_1
\bar \Phi_1)^2 \ ,
\label{anom}
\end{equation}
where
\begin{eqnarray}
\Psi_2 &= &-{m(r+ia\cos\theta)-q^2\over(r-ia\cos\theta)^3
(r+ia\cos\theta)}\, , \nonumber\\ \nonumber\\ \Phi_1 &=& {q \over
\sqrt
2 ( r - i a \cos \theta )^2}\ .
\end{eqnarray}
We have checked that the function (\ref{anom}) does not vanish for any
KN solution with arbitrary values of $m, q, a$ and in particular for
$m=|q|$.  As an additional consistency check we have calculated the
integrated form of the eq.  (\ref{anom}) for Reissner-Nordstr\"om black
hole with $m\geq|q|,\, a=0$.  The result is 2, which agrees with the
well-known Euler characteristic of the Schwarzschild and Kerr black
holes:
\begin{equation}
\chi =  {1\over 32 \pi^2}\int R^{ab} \wedge
R^{cd} \; \epsilon_{abcd}= 2 \ .
\end{equation}

How does this affect the conclusion of the previous section?  Let's
consider now {\it semiclassical} configurations of this theory, that is,
configurations which satisfy the semiclassical equations of motion
obtained by adding first-order quantum corrections to the right-hand
side of the classical equations of motion.  These semiclassical
configurations, then, satisfy the equations (\ref{eqmo}) where the trace
of $J_{\mu\nu}$ is identified with the trace anomaly.  This is indeed a
very small correction which should not produce big changes in the metric
of classical configurations.  In particular, it is reasonable to expect
that classical configurations with nonvanishing imaginary part of $V$
(as the $m=|q|$ KN configurations) will continue to have a nonvanishing
imaginary part of $V$ after the quantum corrections have been taken into
account.

The presence of $J_{\mu}{}^{\mu}\neq 0$ and complex $V$ is incompatible
with the supersymmetry consistency conditions.  Thus, when we embed the
Einstein-Maxwell theory in a supersymmetric theory for which $A\neq 0$
(i.e. the trace anomaly does not vanish) the semiclassical KN
configurations (now including those with naked singularities $m=|q|$)
are not supersymmetric anymore.

The question arises immediately how to make the anomaly coefficient $A$
to vanish.  Looking on the table we may observe that the anomaly
vanishes for any theory which is build out of the $N=4$ multiplet of
supergravity and arbitrary number of $N=4$ Yang-Mills multiplets.  If we
do not want to increase the number of supersymmetries, we may add to
$N=2$ supergravity $11+n$ $N=2$ vector multiplets and $n$
hypermultiplets.  The anomaly of such system of fields vanishes.  What
happens, however, with our naked singularity solutions?

We have found that for all above mentioned theories where the anomaly is
cancelled, supersymmetric configurations with naked singularities are
not solutions of the classical equations of motion anymore.  The
simplest explanation of this mechanism can be given for the $N=4$
theory.  We would like to add new fields to the theory in such a way
that they propagate in the loop diagrams and cancel the anomaly.
Simultaneously we want to make only minimal changes in classical field
equations, in order to preserve our previous solutions.  This is not
possible.  Indeed, in the $N=4$ case there is one new equation for the
dilaton field of the form
\begin{equation}
\nabla^{2}\phi - {\textstyle\frac{1}{2}}
 e^{-2\phi}F^{2} =0 \ .
\label{dil}
\end{equation}
The configurations with naked singularities which we have considered
before had a constant (space- and time-independent) value of the dilaton
field and a non-vanishing value of $F^{2}$.  Thus they do not satisfy
equation (\ref{dil}).  In other words, by adding new fields which cancel
the anomaly, we are adding new equations which are not satisfied by our
old solutions with naked singularities.  This effect is a consequence of
the general structure of the supersymmetric coupling of matter
multiplets to vector fields in gravitational multiplets \cite{toine}.
In particular, the coupling of $N=2$ matter multiplets (and we need at
least 11 vector multiplets to cancel the anomaly) will also result in
additional equations of the type (\ref{dil}) which will invalidate the
naked singularity solutions.

Thus, we have found that in the theory under consideration there are no
stationary supersymmetric solutions with naked singularities.  In the
case of {\it static} solutions studied in \cite{US} this was enough to
show that for nonsupersymmetric configurations the singularities are
even deeper hidden by the horizon, which means that supersymmetry works
as a cosmic censor.  It remains to be seen whether an analogous
statement is true for general stationary solutions.  In any case, the
results obtained above confirm that there exists some deep and
previously unexplored relation between the absence of naked
singularities and supersymmetry.

We are grateful to B. Carter, G. W. Gibbons, M. Grisaru, S. Hawking, G.
Horowitz, A. Linde, P. van Nieuwenhuizen and M. J. Perry for extremely
useful discussions.  We would like to express our gratitude to the
organizers of the programme ``Geometry and Gravity" at the Newton
Institute for the most stimulating atmosphere for work and for the
financial support.  The work of R. K. was also supported by NSF grant
PHY-8612280 and the work of T. O. was supported by European Communities
Human Capital and Mobility programme grant.

\end{document}